\documentclass{aa}
\usepackage{epsfig}
\usepackage{graphics}

\newcommand{\be}{\begin{equation}}
\newcommand{\ee}{\end{equation}}

\begin{document}

\title{Dark energy domination in the Virgocentric flow}

\author{A. D. Chernin\inst{1,2}  \and I. D. Karachentsev\inst{3} \and O.G. Nasonova\inst{3}
\and P. Teerikorpi\inst{1} \and M. J.
Valtonen\inst{1}  \and V. P. Dolgachev\inst{2} \and \\ L. M.
Domozhilova\inst{2} \and G. G. Byrd\inst{4}}

\institute{Tuorla Observatory, Department of Physics and Astronomy, University of Turku, 21500 Piikki\"{o},
Finland \and Sternberg Astronomical Institute, Moscow
University, Moscow, 119899, Russia  \and Special Astrophysical Observatory, Nizhnii Arkhys, 369167,
Russia
\and University of Alabama,
Tuscaloosa, AL 35487-0324, USA }

\authorrunning{A.D. Chernin et al.}
\titlerunning{Dark energy in the Virgocentric flow}

\date{Received / Accepted}

\abstract
{The standard $\Lambda$CDM cosmological model
implies that
all celestial bodies are embedded in a perfectly uniform
dark energy background, represented by Einstein's cosmological constant,
and experience its repulsive antigravity action.
}
{Can dark energy have strong dynamical effects on small
cosmic scales as well as globally? Continuing our
efforts to clarify this question, we focus now on the Virgo Cluster
and the flow of expansion around it.
}
{We interpret the Hubble diagram, from
a new database of velocities and
distances of galaxies in the cluster and its environment, using
a nonlinear analytical model which incorporates the
antigravity force in terms of Newtonian mechanics. The key
parameter is the zero-gravity radius,
the distance at which gravity and
antigravity are in balance.
}
{1. The interplay between the gravity of
the cluster and the antigravity of the dark energy
background determines the kinematical structure of the
system and controls its evolution. 2. The gravity dominates the
quasi-stationary bound cluster, while the antigravity controls the
Virgocentric flow, bringing order and
regularity to the flow, which
reaches linearity and the global Hubble rate
 at distances $\ga 15$ Mpc.
3. The cluster and the flow form a system similar to the
Local Group and its outflow. In the velocity-distance
diagram, the cluster-flow structure reproduces the group-flow
structure with a scaling factor of about 10; the
zero-gravity radius for the cluster system is also 10
times larger.
}  
{The phase and dynamical similarity of the
systems on the scales of 1-30 Mpc suggests that a two-component
pattern may be universal for groups and
clusters: a quasi-stationary bound central component
and an expanding outflow around it, due to the nonlinear gravity-antigravity interplay with the
dark energy dominating in the flow component.}

\keywords{galaxies: Local Group, Virgo cluster, Cosmology: dark matter, dark energy}

\maketitle

\section{Introduction}

Early studies of the motion of our Galaxy (reviewed by Huchra 1988)
led to the discovery of the peculiar velocity of
the Local Group towards the Virgo cluster.
To illustrate,
take the distance to Virgo to be 17 Mpc
and the
Hubble constant $H_0 = 72$ km/s/Mpc, then the expected
cosmological velocity of the cluster is 1224
km/s. The observed velocity is about 1000
km/s. The difference reflects our peculiar velocity of about 220 km/s
towards Virgo.

This estimate of the retarded expansion
(the so-called "Virgo infall") views the universal
expansion as existing not only on "truly
cosmological" scales $\sim 1000$ Mpc, but also on local
scales of $\sim 10$ Mpc only weakly disturbed.
Observational and theoretical aspects of
the Virgo infall were studied e.g. by Silk (1974, 1977), Peebles
(1977), Hoffman \& Salpeter (1982), Sandage (1986), Teerikorpi et
al. (1992), and Ekholm et al. (1999, 2000).

\subsection{The Virgocentric flow}

The Virgo infall is a particular feature of what is
known as the Virgocentric flow:
hundreds of galaxies (our Galaxy among them)
are receeding away from the Virgo cluster. Recent
data (Karachentsev \&
Nasonova 2010) show that the flow velocities range from
nearly zero around 5-8 Mpc from the cluster
center to about 2 000 km/s at the distances of about 30 Mpc.
For $R > 10-15$ Mpc, a roughly
linear velocity-distance relation may be seen in the flow,
making it resemble the global Hubble
expansion. But strong deviations from the linear relation exist
at smaller distances, not surprisingly: 1) the
linear expansion flow, $V = H R$, is the property of the
universe beyond the cosmic "cell of uniformity" ($\ge 300$ Mpc); 2) the Virgo cluster is
a strong local overdensity; 3) the matter
distribution around the cluster is highly nonuniform 
up to the distance of, say, 30 Mpc.

Besides the
natural deviations,
we have here an example of a paradox realized
by Sandage (1986, 1999). He saw a
mystery in the fact that the local expansion proceeds in a quite
regular way.
Moreover, the rate of expansion is similar, if not
identical, to the global
Hubble constant (Ekholm et al. 2001; Thim et al. 2003;
Karachentsev 2005; Karachentsev et al. 2002, 2003b; Whiting 2005).
Originally, the Friedmann theory described the
expansion of a smooth, uniform self-gravitating medium. But
the linear cosmological expansion was discovered by
Hubble where it should not be: in the lumpy environment at distances
$< 20$ Mpc.

\subsection{The local relevance of dark energy}

Soon after the discovery of dark  energy (Riess et al. 1998; Perlmutter et
al. 1999) on global scales,
we suggested (Chernin et al. 2000,
Chernin 2001, Baryshev et al. 2001, Karachentsev et al. 2003a,
Teerikorpi et al. 2005, Chernin et al. 2004) that dark energy with
its omnipresent and uniform density (represented by
Einstein's cosmological constant) may provide the dynamical
background for a regular quiescent expansion on local scales,
resolving the Hubble-Sandage paradox. 
Our key argument came from the fact that the antigravity produced by dark
energy is stronger than the gravity of the Local Group at
distances larger than about 1.5 Mpc from the group center.\footnote{The
increasing observational evidence and theoretical considerations
favoring this view have been discussed by Maccio et al. (2005), Sandage (2006),
Teerikorpi et al. (2006, 2008), Chernin et al. (2006, 2007a,b,c), Byrd et al. (2007), Valtonen et al. (2008),
Balaguera-Antolinez et al. (2007), Bambi (2007), Chernin (2008),
Niemi \& Valtonen (2009), Guo \& Shan (2009).
For some counter-arguments to this new approach see
Hoffman et al. (2008) and Martinez-Valquero et al. (2009).}

The present paper extends our work from
groups to the scale of clusters, focusing on the Virgo cluster.

\section{The phase space structure}

Recent observations of the Virgo cluster and its vicinity 
permit
a better understanding of the physics behind the visible structure and
kinematics of the system.

\subsection{The dataset}

The most complete list of data on the Virgo Cluster
and the Virgocentric flow has been
collected by Karachentsev \& Nasonova (2010). These
include distance moduli of galaxies from the Catalogue of the
Neighbouring Galaxies (= CNG, Karachentsev et al. 2005) and also
from the literature with the best
measurements prefered. Distances from the Tip of the Red Giant Branch (TRGB) and
the Cepheids are used from the CNG together with new TRGB
distances (Karachentsev et al. 2006,
Tully et al. 2006, Mei et al. 2007). For galaxy images in two or
more photometric bands obtained with WFPC2 or ACS cameras at the
HST, the TRGB method yields distances with an accuracy
of about 7\% (Rizzi et al. 2007). The database includes also data
on 300 E and S0 galaxies from the Surface Brightness
Fluctuation (SBF) method by Tonry et al. (2000) with a typical
distance error of 12\%. 
The total
sample contains the velocities and distances of 1371
galaxies within $30$ Mpc from the Virgo
cluster center. Especially interesting is the sample of 761 galaxies
selected to avoid the effect of unknown tangential (to
the line of sight) velocity components. The
velocity-distance diagram for this sample taken from
Karachentsev \& Nasonova (2010) is given in Fig.1.

\begin{figure}
\epsfig{file=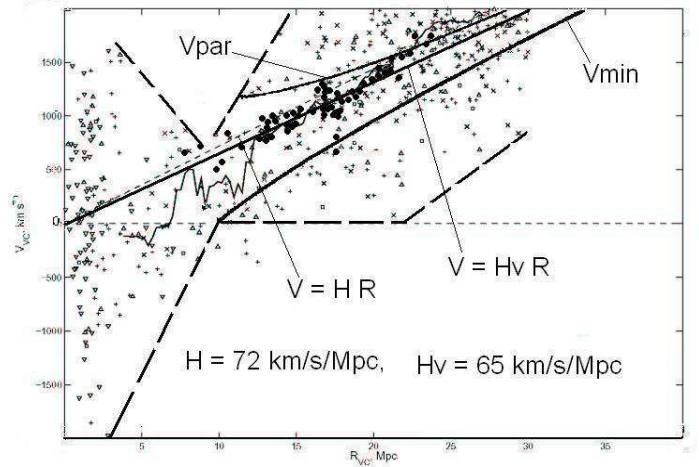, angle=0, width=10.5cm}
\caption{The Hubble diagram with Virgocentric velocities and distances
for 761 galaxies of the Virgo Cluster
and the outflow 
(Karachentsev \& Nasonova 2010). The two-component
pattern is outlined by bold dashed lines.
The zero-gravity radius $R_{\rm ZG} \sim 10 $ Mpc
is located in the
zone between the components.
The
broken line is the running median used by K \&
N to find the zero-velocity radius $R_0 = 6-7$ Mpc. Two
lines from the origo indicate
the Hubble ratios $H = 72$ km/s/Mpc and $H_V =
65$ km/s/Mpc. 
Two curves show special trajectories, corresponding to the
parabolic motion ($E = 0$; the upper one) and
to the minimal escape velocity from the cluster
potential well (the lower one). The flow galaxies with the most accurate distances and
velocities (big dots) occupy the area between these curves.
}
\label{fig1}
\end{figure}

\subsection{The zero-velocity radius and the cluster mass}

The zero-velocity radius within the retarded expansion field around a point-like mass concentration
means the distance where the radial velocity relative to the
concentration is zero. In the ideal case of the mass concentration at rest within
the expanding Friedmann universe this is the distance where  the radial peculiar velocity
towards the concentration
is equal to the Hubble velocity for the same distance.
Using
Tully-Fisher distances in the Hubble diagram, Teerikorpi et al.
(1992) could for the first time see the location of the zero-velocity radius
$R_0$ for the Virgo system, so that $R_0/R_{\rm Virgo} \approx 0.45$ or
$R_0 \approx 7.4$ Mpc.
The work by
Karachentsev \& Nasonova (2010) puts the zero-velocity radius
at $R_0 = 5.0 - 7.5$ Mpc.
For $R < R_0$, positive and negative velocities appear
in practically equal numbers; for $R >
R_0$, the velocities are positive with a few exceptions
likely due to errors in distances.
The zero-velocity radius gives the upper limit of the size of
the gravitationally bound cluster, and the diagram shows that the
Virgocentric flow starts at $R \ge R_0$ and extends at least up to
30 Mpc.

The zero-velocity radius has been often used for estimating the total mass
$M_0$ of a gravitationally bound system. According to
Lynden-Bell (1981) and Sandage (1986), the spherical model with
$\Lambda =0$ leads to the estimator
\be M_0 = (\pi^2/8G) t_U^{-2} R_0^3.\ee
With the age of the universe $t_U = 13.7$ Gyr (Spergel et al. 2007),
Karachentsev \& Nasonova
(2010) find the Virgo cluster mass $M_0 = (6.3 \pm 2.0) \times
10^{14} M_{\odot}$. This result agrees with the virial mass
$M_{\rm vir} = 6 \times 10^{14} M_{\odot}$ estimated by
Hoffman \& Salpeter (1982) and $ \sim 4 \times 10^{14} M_{\odot}$
of Valtonen et al. (1985) and Saarinen \& Valtonen (1985).
Teerikorpi et al. (1992) and Ekholm et
al. (1999, 2000) found that the real cluster mass
might be from 1 to 2 the virial mass, or $(0.6-1.2) \times
10^{15} M_{\odot}$. Tully \& Mohayaee (2004) derived
 the mass $1.2 \times 10^{15} M_{\odot}$ with the
"numerical action" method.

\subsection{The Virgo infall}

Figure 1 shows clearly the Virgo infall:
at distances $R > R_0$, the galaxies
are located mostly below the Hubble line $V_{\rm H} = H
R$ with $H = 72$ km/s/Mpc (Spergel et al. 2007). The deviation of
the median velocity of the flow $V_{\rm m}$ from the Hubble velocity is
well seen in the distance range 7 -- 13 Mpc. At
larger distances the retardation effect gets weaker and
gradually sinks below the measurement error level.

For the distance and peculiar velocity of the Milky Way, we take 
figures from
Karachentsev \& Nasonova (2010): the distance to
the Virgo cluster $17$ Mpc, the recession velocity
$1004 \pm 70$ km/s; with the Hubble constant $H = 72$ km/s/Mpc
the regular expansion velocity is $1224$ km/s. Then the peculiar
velocity directed to the cluster center $\delta V = 220 \pm 70$
km/s (this agrees with the result based on TF distances
and a different analysis by Theureau et al. 1997;
a complete velocity field on the scale of 10--20 Mpc is
given by Tully et al. 2008.)
The effect is not very strong.
It is much more
instructive, however, that there are big nonlinear deviations from
the linear velocity-distance law near the cluster: the
infall effect is strongest there.

\subsection{Two-component phase structure}

The  Virgo system reveals an obvious two-component structure in
Fig.1: the cluster and the
flow, outlined roughly with bold dashed lines.
As we noted, the positive and negative velocities (from -2000 to +1700 km/s)
are seen in the cluster area in equal numbers, so the component is
rather symmetrical relative to the horizontal line $V = 0$.
The border
zone between the components, in the range $6 < R < 12$
Mpc, is poorly populated, and the velocities are
considerably less scattered here, from $-400$ to $+600$
km/s. This zone contains the zero-velocity radius.
The flow component with positive velocities
is at distances $ > 12$ Mpc. It is rather symmetrical
relative to the line $V = H R$. The
velocities are scattered within $\pm 1000$ km/s around
the symmetry line.

A much smaller velocity dispersion is seen in the more accurate data on 75 galaxies
with TRGB and Cepheid distances (mostly from CNG)
located from 10 to 25 Mpc from the cluster center. In the
diagram of Fig.1, the subsample (dots) occupies a strip less
than 500 km/s wide. They are scattered with the mean dispersion of
about 250 km/s around the line $V = H_V R$, where $H_V \simeq 65$
km/s/Mpc. This line is the median line for the subsample.

\subsection{Similarity with the Local system}

The two-component phase structure of the Virgo system looks very
similar to that of the Local system (the Local
Group and the outflow). The diagram of Fig.2 (Karachentsev et
al. 2009) is based on the most complete and accurate data obtained
with the TRGB method using the HST.
The velocities within the Local Group range from $-150$ to $+160$ km/s,
about 10 times less than the velocity spread in
the Virgo cluster. The zone between the two components
of the Local system is in the range $0.7 - 1.2$ Mpc,
also 10 times less than in Fig.1. The local
flow has a dispersion of about 30 km/s around the line $V = H_L
R$, where $H_L = H_V \simeq 60$ km/s/Mpc. This line defines the median
for the local flow. The dispersion is about 10 times less
than in the Vigocentric flow for the subsample of the accurate
distances. The local phase diagram of Fig.2, if
zoomed by a factor of 10, would roughly reproduce the phase structure of
the Virgo system. We will see that the phase
similarity is also supported by the
dynamical similarity.

\begin{figure}
\epsfig{file=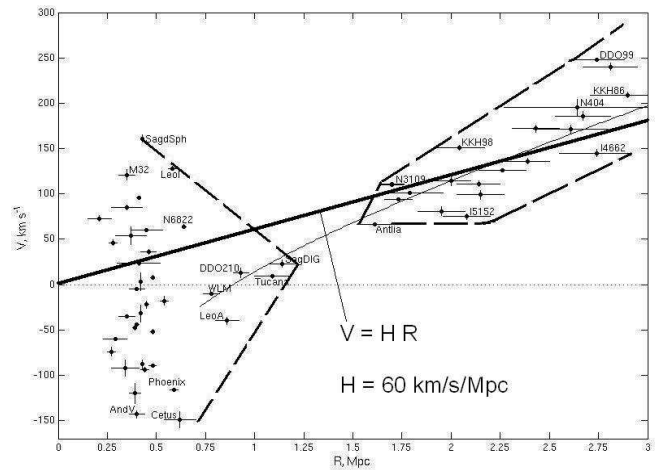, angle=0, width=10cm}
\caption{The Hubble diagram for 57 galaxies of the Local Group and
the local flow (Karachetsev et al. 2009). The thin line indicates
the zero-velocity radius $R_0 = 0.7-0.9$ Mpc. The two-component
phase structure is outlined by the bold dashed line.
Zoomed in about 10 times,
the group-scale structure reproduces well the cluster-scale structure (Fig.1).
The zero-gravity
radius $R_{\rm ZG}$ is about 10 times smaller for the group than
the same key quantity of the cluster system.
The similarity
supports a universal cosmic two-component
grand-design for groups and clusters.
}
\label{fig2}
\end{figure}

\section{Gravity-antigravity interplay}

Dark energy is a form of cosmic energy that produces
antigravity and causes the global cosmological acceleration
discovered by Riess et al. (1998) and Perlmutter et al. (1999) in
observations of SNe type Ia at horizon-size distances of about 1000 Mpc.
These and other observations, especially the cosmic
microwave background (CMB) anisotropies (Spergel et al. 2007),
indicate that the global dark energy density, $\rho_V = (0.73 \pm
0.03)\times 10^{-29}$ g/cm$^3$, makes nearly 3/4 of the
total energy of the universe.

\subsection{Antigravity in Newtonian description}

According to the simplest, most straightforward and quite likely
view adopted in the $\Lambda$CDM cosmology,
dark energy is represented by Einstein's cosmological constant
$\Lambda$, so that $\rho_V = (c^2/8 \pi G) \Lambda$.
If so, then dark energy is
the energy of the cosmic vacuum (Gliner 1965) and is described
macroscopically as a perfectly uniform fluid with the equation of
state $p_V = - \rho_V$ (here $p_V$ is the dark energy pressure;
$c = 1$). This interpretation implies
that dark energy
exists everywhere in space with the
same density and pressure.

Local dynamical effects of dark energy have been studied by
Chernin et al. (2000, 2009), Chernin (2001), Baryshev et al.
(2001), Karachentsev et al. (2003), Byrd et al. (2007), and Teerikorpi et al.
(2008).
For the Local Group and the
outflow close to it,
we showed that the antigravity produced
by the dark energy background exceeds the gravity
of the group at
distances beyond $\simeq 1.5$ Mpc from the group
barycenter. 
Similar results
were obtained for the nearby
M81 and Cen A groups (Chernin et al. 2007a,b).

Extending now our approach to the cluster scale, we consider
the Virgo system (the cluster and the flow) embedded in the
uniform dark energy background. We describe the
system -- in the first approximation -- in a
framework of a spherical model (Silk 1974, 1977; Peebles 1976) where
the cluster is considered as a spherical
gravitationally bound quasi-stationary system. The galaxy
distribution in the flow is represented by a continuous dust-like
(pressureless) matter. The flow is expanding, and it is assumed
that the concentric shells of matter do not intersect each other
in their motion (the mass within each shell keeps constant).

Since the velocities in the cluster and the flow are much less than the speed of light,
we may analyze the dynamics of the cluster-flow system in terms of Newtonian forces.
A spherical shell with a Lagrangian coordinate $r$ and a Eulerian
distance $R(r,t)$ from the cluster center is affected by the
gravitation of the cluster and the flow galaxies
within the sphere of the radius $R(r,t)$. According to the
Newtonian gravity law, the shell experiences an acceleration
\be F_{\rm N} (r,t) = - \frac{GM(r)}{R(r,t)^2},\ee
\noindent in the reference frame related to the cluster center.
Here $M(r)$ is the total mass within the radius $R(r,t)$. 
    
The flow is affected also by the repulsion
produced by the local dark energy. The resulting
acceleration in Newtonian terms is given by the
"Einstein antigravity law":
\be F_{\rm E}(r,t)   = + G\, 2 \rho_{\Lambda} \frac{\frac{4
\pi}{3}R(r,t)^3}{R(r,t)^2} = + \frac{8 \pi}{3} G\rho_{\Lambda}
R(r,t)
\ee
\noindent Here $-2 \rho_{\Lambda} = \rho_{\Lambda} + 3
p_{\Lambda}$ is the effective (General Relativity) gravitating
density of dark energy (for details see, e.g., Chernin
2001). It is negative, and therefore
the acceleration is positive, speeding up the shell
away from the center.

\subsection{Zero-gravity radius in the point mass approximation}

The equation of motion of an individual shell has the form
\be \ddot R(r,t) = F_{\rm N} + F_{\rm E} = - \frac{GM(r)}{R(r,t)^2} + \frac{8 \pi}{3}
G\rho_{\Lambda} R(r,t). \ee
\noindent Its integration leads to the mechanical energy conservation law
for each matter shell:
\be \frac{1}{2} V(r,t)^2 = \frac{G M(r)}{R(r,t)} + \frac{4 \pi}{3}
G\rho_{\Lambda} R(r,t)^2 + E(r). \ee
Here $V(r,t) = \dot R(r,t)$ is the Eulerian radial
velocity of the shell, and $E(r)$ is its total mechanical energy
(per unit mass). This is the basic equation which describes the
outflow in our model. It is the Newtonian
analogue of the 
LTB
spacetime of General
Relativity (e.g., Gromov et al. 2001).

A simpler model results, if we can assume that
the cluster mass $M_0$ makes the majority of the mass
$M(r)$, and the flow galaxies can be treated as test particles.
Curiously enough, this assumption makes the model more general in
the sense that no spherical symmetry is required now for the
galaxy distribution in the flow (not necessarily
concentric shells around the cluster). However, the gravity force
produced by the
cluster at the distances of the flow is still isotropic. The model remains also
nonlinear: no restriction on the strength of the dynamical effects
is assumed.

The mass-point model may seem overly simple, as it is known that
the Virgo cluster contributes
only 20 \% of the bright galaxies within the volume up to the distance
of the Local Group (Tully 1982). However, Tully \& Mohayaee (2004)
conclude from their numerical action method that the mass-to-luminosity
ratio of the E-galaxy rich Virgo cluster must be several times higher (they give
$M/L \sim 900$) than in the surrounding region ($\sim 125$)
in order to produce the infall velocities observed
close to Virgo (a qualitatively similar result was found by Teerikorpi et al. 1992).
 Therefore, the model with most of the mass
in and close to the Virgo cluster
could be a sufficient first approximation. 

In this simplified/generalized version, the mass $M(r) = M$ is
constant in Eq. 4, which shows that antigravity and
gravity balance each other at the distance $R = R_{\rm ZG}$, where
\be R_{\rm ZG} = [\frac{3 M}{8\pi \rho_{\Lambda}}]^{1/3} \ee

\noindent is the zero-gravity radius (Chernin et al. 2000; Chernin
2001; Dolgachev et al. 2003,2004). Gravity dominates at $R <
R_{\rm ZG}$, and antigravity is stronger than gravity at $R > R_{\rm ZG}$.
The zero-gravity radius is the major quantitative factor
in the gravity-antigravity interplay in the system.

According to various estimates (Sect.2),
the total mass
of the Virgo cluster is 1 -- 2 times the
virial mass: $M = (0.6-1.2) \times 10^{15} M_{\odot}$.
Then the known global $\rho_{\Lambda}$ gives
\be R_{\rm ZG} = (9-11) \;\;\; {\rm Mpc}.\ee
Karachentsev \& Nasonova (2010) argue in favour of
one virial mass for the cluster mass; if so
the lower limit for $R_{\rm ZG}$ in Eq.(7) may be
more realistic.

Eqs.6,7 indicate that the zero-gravity radius $R_{\rm ZG}$ of the
Virgo system is not far from
the zero-velocity radius $R_0 =
5.0 - 7.5$ Mpc.
In fact, one expects that $R_{\rm ZG}$ is slightly larger
than $R_0$ (Teerikorpi et al. 2008) in the standard cosmology.
The zero-velocity radius constrains the size of the cluster
(Sec.2), and there also gravity dominates ($R < R_0 \la
R_{\rm ZG}$), the necessary condition for a gravitationally bound cluster.
On the other hand, the Virgocentric
flow is located beyond the distance $R_{\rm ZG} \simeq R_0$, hence
the flow is dominated dynamically by the dark energy
antigravity.

The zero-gravity radius of the Virgo system is about 10
times larger than that of the Local system for which
$R_{\rm ZG} = 1-1.3$ Mpc (Chernin 2001; Chernin et al. 2000, 2009). We
see here the same factor 10 as for the
phase structures of the Virgo and Local systems (Sec.2). Thus
the phase similarity is complemented by the dynamical
similarity.

\subsection{Antigravity push}

Now we return to the Milky Way and estimate the net gravity-antigravity
effect at its distance $R_{\rm MW} = 17$ Mpc from the
cluster center. With $R_{\rm ZG} = 9-11$ Mpc, we find that here
 the antigravity force is stronger than the gravity:
\be F_{\rm E}/|F_{\rm N}| \simeq 4-7, \;\;\; R = R_{\rm MW}.\ee
As we see, the extra gravity produced by the Virgo
cluster mass is actually overbalanced by the antigravity of the
dark energy located within the sphere of the radius $R_{\rm MW}$
around the cluster. Thus the Milky Way experiences not a
weak extra pull (cf. Secs.1,2), but rather a strong push (in the
Virgocentric reference frame) from the Virgo direction.

The dark energy domination in the flow changes the current dynamical
situation drastically in comparison with the earlier models by
Silk (1974, 1977) and Peebles (1976). At the periphery of the
Virgocentric flow (at about 30 Mpc) the force ratio is
impressively large: $F_E/|F_N| \simeq 20-37$.

\subsection{Antigravity domination}

The basic trend of the flow evolution is seen
from Eq.5. With $M(r) = M = {\rm const}$, it takes at
large times the form:
\be V(r,t = \infty) = [\frac{8 \pi}{3} G\rho_{\Lambda}]^{1/2} R.
\ee
\noindent In this limit, $F_{\rm E}/|F_{\rm N}| \rightarrow
\infty$, and the gravity of the cluster is practically negligible
compared to the dark energy antigravity. As a
result, the flow acquires asymptotically the
linear velocity-distance relation with the constant Hubble factor
expressed via the dark energy density only:
\be H_{\Lambda} =  [\frac{8 \pi G}{3}\rho_{\Lambda}]^{1/2} = 63-65
\;\;\; {\rm km/s/Mpc}, \ee
\noindent At a given moment of time (that of observation) $t = t_0$
Eq.5  presents the spatial structure of the flow.
 At large
distances
\be V(r = \infty,t_0) \rightarrow H_{\Lambda}R(r,t_0).\ee
The flow trajectories converge to the asymptotic
of Eq.11 independently, in general, of the "boundary/initial
conditions" near the cluster (see figure 2 in Teerikorpi \& Chernin
2010). In particular,  strong
perturbations at small distances in the flow are
compatible with its  regularity at larger distances.
No tuning of
the model parameters are required for this -- the only
significant
quantity is the zero-gravity radius $R_{\rm ZG}$
determined by the cluster mass and the DE
density.

Thus, dark energy dominates the dynamics of the
Virgocentric flow on its entire observed extension from 12 to at least 30
Mpc. Because of the strong antigravity, the flow proves to
be rather regular acquiring the linear Hubble relation
at distances $\ga 15$ Mpc from the cluster center. This
major conclusion from the model also resolves
the Hubble-Sandage paradox on the scale of
$\sim 10$ Mpc, in harmony with our general
approach to the old puzzle (Sec.1).

\section{Discussion}

We discuss here some specific features and implications of
the dark energy domination in the Virgocentric flow.

\subsection{The Einstein-Straus radius}

The mass of dark matter and baryons in
the Virgo Cluster, $M =
(0.6-1.2)\times 10^{15} M_{\odot}$, was initially collected from
the volume which at the present epoch has the radius $
R_{\rm ES} = [\frac{3M}{4\pi \rho_M(t_0)}]^{1/3} = 15-19$
Mpc, where $\rho_M(t_0)$ is the mean matter density in
the Universe. The quantity $R_{\rm ES}$ is known as the
Einstein-Straus radius in the vacuole model. In the presence of
dark energy (Chernin et al. 2005; Teerikorpi et al. 2008), $R_{\rm ES}
= [\frac{2 \rho_{\Lambda}}{\rho_M(t_0)}]^{1/3} R_{\rm ZG} = 1.7
R_{\rm ZG}$.
Within the  E-S vacuole, the
spherical layer between the radii $R_0$ and $R_{\rm ES}$ does not contain
any mass. Therefore in this layer, the approximation of test
particles adopted in the model of Sec.3 works well. Beyond the
distance $R_{\rm ES}\simeq R_{\rm MW}$, the model needs modification,
but remains qualitatively correct up to the distances of 25-30
Mpc. Anyway since $2\rho_{\Lambda} > \rho_M(t_0)$, the antigravity
of dark energy dominates in the entire area of the Virgocentric
flow in Fig.1.

\subsection{Two special trajectories}

Eq.5 shows that the flow has zero velocity at $R = R_{\rm ZG} \simeq
R_0$, if the total energy of the shell or an individual
galaxy
\be E(r) = E_{\rm min} = -\frac{3}{2} \frac{GM}{R_{\rm ZG}}. \ee It is
easy to see that $E_{\rm min}$ is the minimal energy needed for a
particle to escape from the potential well of the cluster. Eq.5
with $E(r) = E_{\rm min}$ gives the corresponding velocity
$V_{\rm min}(R)$ of the flow at different distances in the flow:
\be V_{\rm min} = H_{\Lambda} R [1 + 2 (\frac{R_{\rm ZG}}{R})^3 - 3
(\frac{R_{\rm ZG}}{R})^2]^{1/2}.\ee

\noindent At large distances ($R >> R_{\rm ZG}$), the trajectories
with $E(r) = E_{\rm min}$ approach the asymptotic given by Eq.11.

Another special value of the mechanical energy is $E(r) = 0$,
corresponding to parabolic motion. In this case,
\be V_{\rm par} = H_{\Lambda} R [1 + 2 (\frac{R_{\rm ZG}}{R})^3] ^{1/2}.\ee

\noindent The two special trajectories with $E(r) = E_{\rm min}$ and
$E(r) = 0$ (see Fig.1) converge to the asymptotic of Eq.11 at
large distances -- in agreement with the results of Sec.3.

It is quite interesting that the most
accurate data on the flow (from the TRGB and Cepheids;
dots in Fig.1) show a low dispersion around
the line $V = H_{\rm V} R$, and the corresponding
trajectories occupy the area just between the curves $V_{\rm min}(R)$
and $V_{\rm par}(R)$. The rather wide spread of the
other data in Fig.1 is mostly due to observational errors
(Karachentsev \& Nasonova 2010 -- especially
their Fig.7b). Anyway, the restriction $E_{\rm min} \le E \le
0$ puts an upper limit on the value $|E|$ in Eq.5, and because of
this, the asymptotic of Eq.11 may be reached more easily in the
flow.

It is also suggestive that the Local Group outflow has the same feature:
the galaxies prefer the area between $V_{\rm min}$ and
$V_{\rm par}$ in the Hubble diagram of Fig.2 (c.f.
Teerikorpi et al. 2008). This is another similarity aspect
between the Virgo system and the Local system (Sec.2). Mathematically, the
similarity reflects the fact that Eqs.13
and 14 have exactly the same view for both systems in the
dimensionless form $y(x) = 0$ with $y = V/(H_{\Lambda}R_{\rm ZG}), x =
R/R_{\rm ZG}$; the scale factor is again about 10.

This similarity suggests that the condition $E_{\rm min} \le E \le 0$
has more general significance; it may reflect
the kinematical state of the flow at the early formation epoch.
For example, $V > V_{\rm min}$ follows, if
the galaxies of the flow escaped initially from the cluster (Chernin et al. 2004, 2007c).

\subsection{Comparison with cosmology}

The zero-gravity radius $R_{\rm ZG}$ is a local spatial counterpart of
the redshift $z_{\Lambda} \simeq 0.7$ at which the
gravity and the antigravity balance each other in the Universe as
a whole. Globally, the balance takes place
only at one moment $t(z_{\Lambda})$ in the
entire co-moving space, while the local gravity-antigravity
balance exists during all life-time of the system since its
formation, but only on the sphere of the radius $R_{\rm ZG}$.

The Virgocentric flow should not be identified
with the global cosmological expansion. The Virgo
system lies deeply inside the cosmic cell of uniformity and
does not "know" about the universe of the horizon scales.
In the early epoch of weak protogalactic perturbations, the dark
matter and baryons of the system participated in the cosmological
expansion. But later, at the epoch of nonlinear perturbations, the
matter separated from the expansion and underwent violent
evolution. Contrary to that, the unperturbed flow at
horizon-scale distances was not similarly affected
and up to now the initial isotropy of its
expansive motion and the uniformity of its matter distribution are conserved.
Therefore, it is mysterious that the global Hubble
constant $72$ km/s/Mpc is so close to the local
expansion rate $65$ km/s/Mpc in the Virgocentric
flow (two beams in Fig.1). This is one aspect of the
Hubble-Sandage paradox (Sec.1).

Trying to find an explanation, we note that the
Friedmann equation for the scale factor $a(t)$ has
much the same form as the local
energy conservation equation (5):
\be \frac{1}{2} \dot a(t)^2 = \frac{G C}{a(t)} + \frac{4
\pi}{3}G\rho_{\Lambda} a(t)^2 + B. \ee
\noindent The constant $C = \frac{4 \pi}{3}\rho_{M}(t) a(t)^3$ has
the dimension of mass, and $\rho_{M}(t)$ is the
density of matter (dark and baryonic);
the energy constant $B = 0$ in the standard flat
cosmology. The
cosmological Hubble factor comes from this equation:
\be H(t) = \frac{\dot a(t)}{a} = [\frac{8 \pi G}{3}(\rho_{M}(t) +
\rho_{\Lambda})]^{1/2} = H_{\Lambda} [1 + \frac{\rho_m}{2
\rho_{\Lambda}}]^{1/2}. \ee
When time goes to infinity, the matter density drops to
zero, the dark energy becomes dominant, and the cosmological
Hubble factor becomes time-independent:
\be H(t) \rightarrow H_{\Lambda} =  [\frac{8 \pi
G}{3}\rho_{\Lambda}]^{1/2} = 63-65 \;\;\;{\rm km/s/Mpc}. \ee
At the present epoch, $t_U = 13.7$ Gyr, the Hubble
rate
\be H = H_{\Lambda} [1 + \frac{\Omega_M}{ \Omega_{\Lambda}}]^{1/2}
\simeq 72 \;\;\; {\rm km/s/Mpc}, \ee
which is equal to the empirical (WMAP) value.

As we see, the Virgocentric flow and the global cosmological
expansion have a common asymptotic in the limit of dark energy
domination: for the local flow, this is the asymptotic in space
(large distances from the cluster center), and for the global
flow, this is the asymptotic in time (a future of the Universe).
If the asymptotic is approached now both globally and locally, we
would have $H \simeq H_{\Lambda}$ on the largest scales and $H_V
\simeq H_{\Lambda}$ in the Virgocentric flow.

In their present states, both flows are near the
asymptotic, though do not quite reach it. That is why
$H$, $H_V$, and $H_{\Lambda}$ have similar values.
The flows are not related to each other directly;
however, they "feel" the same uniform dark energy
background, and in this way both are controlled by the antigravity
force which is equally effective near the cosmic
horizon and at the periphery of the Virgocentric flow.  As a
result, the local flow at its large distances looks
much like a "part" of the global horizon-distance flow.

The near coincidence of the values $H$, $H_{\rm V}$, and $H_{\Lambda}$
explained by our model clarifies the Hubble-Sandage paradox
(Sec.1) for the Virgocentric flow: asymptotically, the flow
reaches not only the linear Hubble law (Sec.3),
but also acquires the expansion rate $H_V$ which approaches
the global rate when the
measuring accuracy is improved.

\subsection{Probing local dark energy}

The global density of dark energy can be derived, if the
redshift $z_{\Lambda}$ is found at which the gravity of matter is
exactly balanced by the dark energy antigravity, and also the
present matter density is known. The same logic works
in local studies: just find the
zero-gravity radius $R_{\rm ZG}$ from the outflow
observations and the mass of the local system.

We may now
restrict the value of $R_{\rm ZG}$ using the
diagram of Fig.1. Since the zero-gravity surface lies outside the
cluster volume, it should be that $R_{\rm ZG} > R_0 \simeq 6-7$ Mpc.
On the other hand, the linear velocity-distance
relation, with the Hubble ratio close to $H$, is clearly seen from a distance of about 15 Mpc.
This suggests
that $R_{\rm ZG} < 15$ Mpc. This argument that
the zero-gravity surface is located below the point where the local flow
reaches the global expansion rate, gained support from
the calculations by Teerikorpi \& Chernin (2010).
   As for the cluster mass, we adopt the
range $M = (0.6-1.2) \times 10^{15} M_{\odot}$ (Sec.2). Then Eq.6
in the form $\rho_{x} = M/(\frac{8\pi}{3} R_{\rm ZG}^3)$
directly leads to robust upper and lower limits to the unknown
local dark energy density $\rho_x$:
\be 0.2 < \rho_{x} < 5 \times 10^{-29} \;\; {\rm g/cm^3}. \ee \noindent
\noindent In fact, the upper limit obtained from the cluster size ($R_{\rm ZG}$)
is conservative. We note that the theoretical minimum velocity
curve in Fig.1 would be shifted up to the velocity-distance relation
defined by the TRGB and Cepheid distances, if one takes
$\rho_{x} = 1.4 \times 10^{-29} \; {\rm g/cm^3}$. This would be
a more realistic upper limit, and near the limit
similarly obtained in Teerikorpi et al. (2008) using the Local Group and
the M81 group.
Thus
the local dark energy density is near the global value or
may be the same;  $\rho_{\Lambda} = 0.73 \times 10^{-29}$ g/cm$^3$ lies comfortably
in the range of Eq.19.
Similar estimates for the dark energy come from studies of the Local
Group, M81 group and Cen A group (Chernin et al. 2007a,b, 2009).

We may now argue the other way round: let us assume that the
density of dark energy is known from cosmological observations and the
zero-gravity radius $R_{\rm ZG}$ is between 6 and 15 Mpc. Then we may
estimate independently the mass of the Virgo Cluster: $M =
(0.3-4)\times 10^{15} M_{\odot}$. The value of the cluster mass
adopted earlier is within this interval.

\subsection{The origin of the local flows?}

The model of Sec.3 describes the present
structure of the Virgocentric flow and also predicts its
future evolution. However, it says almost nothing about the
initial state of the flow.
Our discussion suggests that the
Virgo cluster and the outflow form a
system with a common origin and evolutionary history. Its
formation was most probably due to complex linear and nonlinear
processes (collisions and merging of galaxies and
protogalactic units, violent relaxation of the system in its
self-gravity field, etc.). Similar processes are expected
in the early history of the Local Group and other systems with the
two-component structure. Their basic features
might be recognized from big
N-body simulations, like the Millennium Simulation
(Li \& White 2008).
We leave it for a later study to clarify the way in which the
galaxies of local flows gain their initial expansion
velocities.

\section{Conclusions}

The recently published systematic and most accurate data on the
velocities and distances of galaxies in the Virgo cluster and its
environment (Karachentsev \& Nasonova 2010) shed light on the
physics behind the observed properties of the cluster and
the flow. We find that:

1. The cluster and the flow can be treated together as a
physical system embedded in the uniform dark energy
background; this is adopted in our analytical nonlinear model.

2. The nonlinear interplay between the gravity produced by the
cluster mass (dark matter + baryons) and the antigravity
of the dark energy  is the major dynamical
factor determining the kinematic structure of the system and
controlling its evolution; our model enables one to describe this in
a simple (exact or approximate) quantitative way.

3. The key physical parameter of the system is its zero-gravity
radius, $R_{\rm ZG} = 9-11$ Mpc,
below which the gravity dominates the quasi-stationary bound cluster, while
the antigravity dominates the expanding Virgocentric flow.

4. The dark energy antigravity brings order and
regularity to the Virgocentric flow and the flow
acquires the linear velocity-distance relation at the
Virgocentric distances $R > 15$ Mpc. Thus, the Hubble-Sandage
paradox is understood on the cluster scale.

5. In the velocity-distance diagram, the
zero-gravity radius is near to or somewhat larger than the zero-velocity
radius $R_0 = 5.0-7.5$ Mpc, as determined by Karachentsev \& Nasonova
(2010).

6. With the value of $R_{\rm ZG} \ga R_0$ found from the diagram,
one may estimate the mass of the Virgo cluster, using the known global density of
dark energy: $M = (0.3-4)\times 10^{15} M_{\odot}$.

7. With the same value of $R_{\rm ZG}$, one may estimate the local
density of the dark energy in the Virgo system, if the mass of the
cluster is considered known (from the virial or the
zero-velocity method): $\rho_{\Lambda} = (0.2-5) \times 10^{-29}
{\rm g/cm}^3.$

8. The two-component Virgo system is similar to the Local system
formed by the Local Group together with the local outflow around
it. In the velocity-distance diagram, the Virgo system
reproduces the Local system with the scale factor of
about 10. A dynamical similarity is seen in
the fact that the zero-gravity radius for the Virgo system is
about 10 times larger than for the Local system.

9. The similarity  suggests that a universal two-component grand
design may exist with a quasi-stationary bound central component
and an expanding outflow around it on the spatial scales of 1-30
Mpc. The phase and dynamical structure of the real two-component
systems reflects the nonlinear gravity-antigravity interplay with
dark energy domination in the flow component.

\acknowledgements
 A.C., V.D., and L.D. thank the RFBR for partial support via the grant 10-02-00178.
We also thank the anonymous referee for useful comments.

{}

\end{document}